\documentclass[%
 reprint,
 superscriptaddress,
 nobibnotes,
 amsmath,amssymb,
 prl,
floatfix,
longbibliography,
]{revtex4-1}
\usepackage[usenames,dvipsnames,svgnames,table]{xcolor}
\colorlet{m}{Black}
\colorlet{k}{Black}

\usepackage{graphicx}
\usepackage[outdir=./]{epstopdf}

\usepackage{dcolumn}
\usepackage{bm}
\usepackage[colorlinks=true,citecolor=blue]{hyperref}

\begin{document}

\title{Broken Symmetry Effects due to Polarization on Resonant Tunneling Transport in Double-Barrier Nitride Heterostructures}
\author{Jimy Encomendero}
\email{jje64@cornell.edu}
\affiliation{School of Electrical and Computer Engineering, Cornell University, Ithaca NY 14853 USA}%

\author{Vladimir Protasenko}
\affiliation{School of Electrical and Computer Engineering, Cornell University, Ithaca NY 14853 USA}%

\author{\mbox{Berardi Sensale-Rodriguez}}
\affiliation{\hbox{Department of Electrical and Computer Engineering, The University of Utah, Salt Lake City UT 84112 USA}}%

\author{Patrick Fay}
\affiliation{Department of Electrical Engineering, University of Notre Dame, Notre Dame IN 46556 USA}%

\author{Farhan Rana}
\affiliation{School of Electrical and Computer Engineering, Cornell University, Ithaca NY 14853 USA}%

\author{Debdeep Jena}
\email{djena@cornell.edu}
\affiliation{School of Electrical and Computer Engineering, Cornell University, Ithaca NY 14853 USA}%
\affiliation{Department of Materials Science and Engineering, Cornell University, Ithaca NY 14853 USA}%

\author{Huili Grace Xing}
\email{grace.xing@cornell.edu}
\affiliation{School of Electrical and Computer Engineering, Cornell University, Ithaca NY 14853 USA}%
\affiliation{Department of Materials Science and Engineering, Cornell University, Ithaca NY 14853 USA}%
\affiliation{\hbox{Kavli Institute at Cornell for Nanoscale Science, Cornell University, Ithaca, New York 14853, USA}}%


\begin{abstract}
\,\newline The phenomenon of resonant tunneling transport through polar double-barrier heterostructures is systematically investigated using a combined experimental and theoretical approach. On the experimental side, GaN/AlN resonant tunneling diodes (RTDs) are grown by molecular beam epitaxy. $\textit{In-situ}$ electron diffraction is used to monitor the number of monolayers incorporated into each tunneling barrier of the RTD active region. Using this precise epitaxial control at the monolayer level, we demonstrate exponential modulation of the resonant tunneling current density as a function of barrier thickness. At the same time, both the peak voltage and characteristic threshold bias exhibit a dependence on barrier thickness as a result of the intense electric fields present within the polar heterostructures. To get further insight into the asymmetric tunneling injection originating from the polar active region, we present an analytical theory for tunneling transport across polar heterostructures. A general expression for the resonant tunneling current which includes contributions from coherent and sequential tunneling processes is introduced. After the application of this theory to the case of GaN/AlN RTDs, their experimental current-voltage characteristics are reproduced over both bias polarities, with tunneling currents spanning several orders of magnitude. This agreement allows us to elucidate the effect of the internal polarization fields on the magnitude of the tunneling current and broadening of the resonant tunneling line shape. Under reverse bias, we identify new tunneling features originating from highly attenuated resonant tunneling phenomena, which are completely captured by our model. The analytical form of our model, provides a simple expression that reveals the connection between the design parameters of a general polar RTD and its current-voltage characteristics. This new theory paves the way for the design of polar resonant tunneling devices exhibiting efficient resonant current injection and enhanced tunneling dynamics as required in various practical applications.
\\

\end{abstract}

\maketitle

\section{I. Introduction}

Resonant tunneling of electrons is an ultra-fast quantum transport process that is essential for the operation of various electronic and photonic devices. The versatility of this transport regime stems from the possibility of tuning the tunneling dynamics by means of bandgap engineering, enabling ultrafast carrier injection into discrete energy levels [\onlinecite{Harada1986,Faist1994,Kanaya2014,Kanaya2015}].

This important feature has been exploited over recent decades for the design of increasingly fast resonant tunneling diodes (RTDs), leading to the demonstration of subpicosecond tunneling times on the order of approximately 35~fs [\onlinecite{Kanaya2015}]. The possibility of engineering transport dynamics, combined with their characteristic negative differential conductance (NDC), makes RTDs attractive the manufacture of ultrafast electronic oscillators [\onlinecite{Izumi2017,Kanaya2014,Kanaya2015}]. With oscillation frequencies well inside the terahertz band, RTDs stand out as viable sources of terahertz radiation used in practical applications such as high-data-rate communication networks and on-chip spectroscopy systems [\onlinecite{Watson2017,Oshima2016,Kitagawa2017}].

On the photonics side, resonant tunneling transport is engineered to control the injection and depopulation times of the upper and lower lasing levels of quantum cascade lasers (QCLs)[\onlinecite{Faist1994}]. In these devices, the characteristic tunneling times, controlled by the height and thickness of the tunneling barriers, are designed to attain population inversion and optical gain. As an added benefit, QCLs offer the possibility of tuning their lasing frequency by means of band-structure engineering. As a result, a broad range of operating frequencies can be obtained using a single material system. This versatility makes QCLs useful in a variety of practical applications, including metrology [\onlinecite{Bartalini2014}], spectroscopy [\onlinecite{Hubers2013}] and biomedical imaging [\onlinecite{Bird2015}].

Over recent decades, considerable progress has been made in the performance of terahertz RTD oscillators and QCLs manufactured with well-developed semiconductor materials, such as GaAs/Al$_x$Ga$_{1-x}$As [\onlinecite{Brown1989}] and In$_x$Ga$_{1-x}$As/In$_y$Al$_{1-y}$As~[\onlinecite{Kanaya2014}]. Despite these advances, those technologies still exhibit limitations that prevent their widespread use. In this scenario, alternative material systems are necessary to overcome the power and temperature limitations of arsenide-based resonant tunneling injection.

The III-nitride family of wide-band-gap semiconductors stands out as a promising alternative for developing resonant tunneling devices. During the last two decades considerable effort has been dedicated to the engineering of resonant carrier injection within this material system [\onlinecite{Kikuchi2001,Kikuchi2002,Foxon2003,Belyaev2004,Hermann2004,Golka2006,Kurakin2006,Petrychuk2007,Bayram2010a,Bayram2010b,Bayram2010c,Songmuang2010,Boucherit2011,Carnavale2012,Bayram2012,Li2012,Li2013,Shao2013,Nagase2014,Nagase2015,Grier2015,Encomendero2016X,Growden2016,Encomendero2017,Growden2018,Encomendero2018}]. At the heart of this initiative, lies the outstanding material properties that III-nitride heterostructures offer for the design of high-power RTD oscillators and room temperature intersubband (ISB) lasers. 

III-nitride semiconductor materials hold the promise for extending the lasing frequencies and operating temperatures of intersubband emitters. This is possible due to the high LO-phonon energy ($\hbar\omega_{\text{LO}}\approx$~92~meV in GaN), which prevents thermal depopulation of the upper lasing level at room temperature. With their large conduction-band offsets and high breakdown electric fields, nitride heterojunctions are also suitable for high-power applications. On the basis of these properties, RTDs manufactured with nitride semiconductors stand out as ideal candidates for high-power ultrafast electronic oscillators.

The first microwave oscillators driven by III-nitride RTDs were recently fabricated by the authors, using GaN/AlN double-barrier heterostructures as gain elements [\onlinecite{Encomendero2018}]. This milestone, in conjunction with the demonstration of high resonant tunneling currents at room temperature [\onlinecite{Encomendero2018,Growden2018}], pave the way for developing III-nitride ultrafast THz oscillators.

Harnessing nitride-based resonant tunneling injection requires however, a thorough understanding of the role played by the strong polarization fields present along the tunneling path. The noncentrosymmetric crystal structure of III-nitride materials gives rise to intense spontaneous and piezoelectric polarization fields, which have been used for band-structure engineering. This technique, known as ``polarization engineering", has been implemented in various electronic and photonic devices to induce two-dimensional and three-dimensional electron gases [\onlinecite{Khan1992,Jena2003}], assist in p-type doping [\onlinecite{Simon2010}], and enhance Zener interband tunneling [\onlinecite{Simon2009}]. However, the important consequences of these internal electric fields on intraband resonant tunneling, have not been systematically investigated. 

Tunneling transport, being exponentially sensitive to potential barriers and electric fields, stands as a unique experimental probe into the dramatic effects of the polarization fields. In a previous report, we elucidated the crucial role played by these internal fields in the main tunneling features of polar III-nitride RTDs [\onlinecite{Encomendero2017}]. The asymmetric tunneling injection, induced by the polar heterostructure, manifests not only in the forward resonant voltage and peak tunneling current, but also in the reverse-bias direction. In this regime, a critical threshold voltage, unique in polar RTDs, was identified for the first time [\onlinecite{Encomendero2017}].

In the present work, on the basis of a combined experimental and theoretical approach, we investigate the control exerted by the tunneling barriers on the resonant current injected through the polar double-barrier structure. Experiments comprise the growth, fabrication, and characterization of GaN/AlN RTDs in which we engineer different magnitudes of resonant current densities by systematically varying the thickness of the tunneling barriers.

On the theoretical side, we develop an analytical transport theory for tunneling injection across polar RTDs. The Landauer-B\"uttiker quantum transport model is employed to derive an expression for the resonant tunneling current which includes contributions from coherent and sequential tunneling processes [\onlinecite{Luryi1985,Stone1985,Buttiker1988,Landauer1987}]. This analytical approach sheds light on the asymmetric tunneling transport caused by the polar active region. After the application of this theory to the case of GaN/AlN RTDs, the experimental current-voltage ($\textit{J-V}$) characteristics are reproduced over several orders of magnitude. This agreement allows us to elucidate the role played by the internal polarization fields on the magnitude of the tunneling current and broadening of the resonant tunneling line shape. Furthermore, under reverse-bias injection, we identify new tunneling features originating from highly attenuated resonant tunneling phenomena, which are also completely captured by our model.

\begin{figure*}[t]
	\centering
	\includegraphics[width=\textwidth]{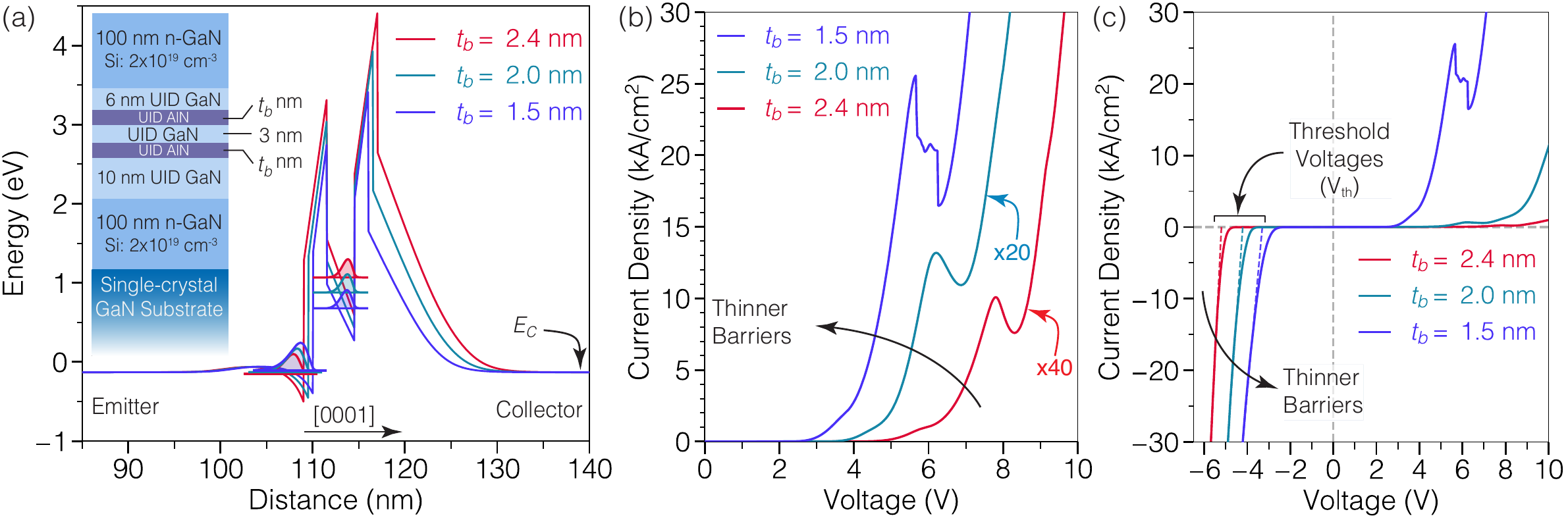}
	\hypertarget{Fig1}{}
	\caption{Different magnitudes of resonant tunneling current are engineered in GaN/AlN RTDs by systematically varying the thickness of the AlN barriers. (a) The equilibrium conduction-band diagram ($E_C$) of the double-barrier active region of each sample is calculated with the use a self-consistent Schr\"odinger-Poisson solver [\onlinecite{Tan1990}]. The energy of the ground-state inside the quantum well raises toward higher values with increasing barrier thickness. The inset schematically depicts the complete device structure, including the unitentionally doped (UID) spacer layers and doping levels of the contact regions. When the devices are biased, asymmetric current-voltage characteristics are observed. (b) Under forward bias, the characteristic resonant peak and negative differential conductance shift towards larger voltages for thicker AlN barriers. Because of the strong tunneling attenuation, the magnitude of the peak current is modulated over two orders of magnitude when the barrier thickness is increased from 1.5~nm to 2.4~nm. The currents in (b) have been scaled by different factors to facilitate direct comparison. (c) Under reverse bias, the polarization-induced threshold voltage ($V_\text{th}$) exhibits a linear dependence on the barrier thickness. This result, consistent with theoretical predictions, allows measurement of the internal polarization fields [\onlinecite{Encomendero2017}]. The magnitude of this critical voltage is experimentally determined using the linear interpolation procedure described in section II. The dashed lines indicate the linear fits, with $V_\text{th}$ as voltage intercept.}
\end{figure*}

\begin{figure*}[t]
	\hypertarget{Fig2}{}
	\centering
	\includegraphics[width=\textwidth]{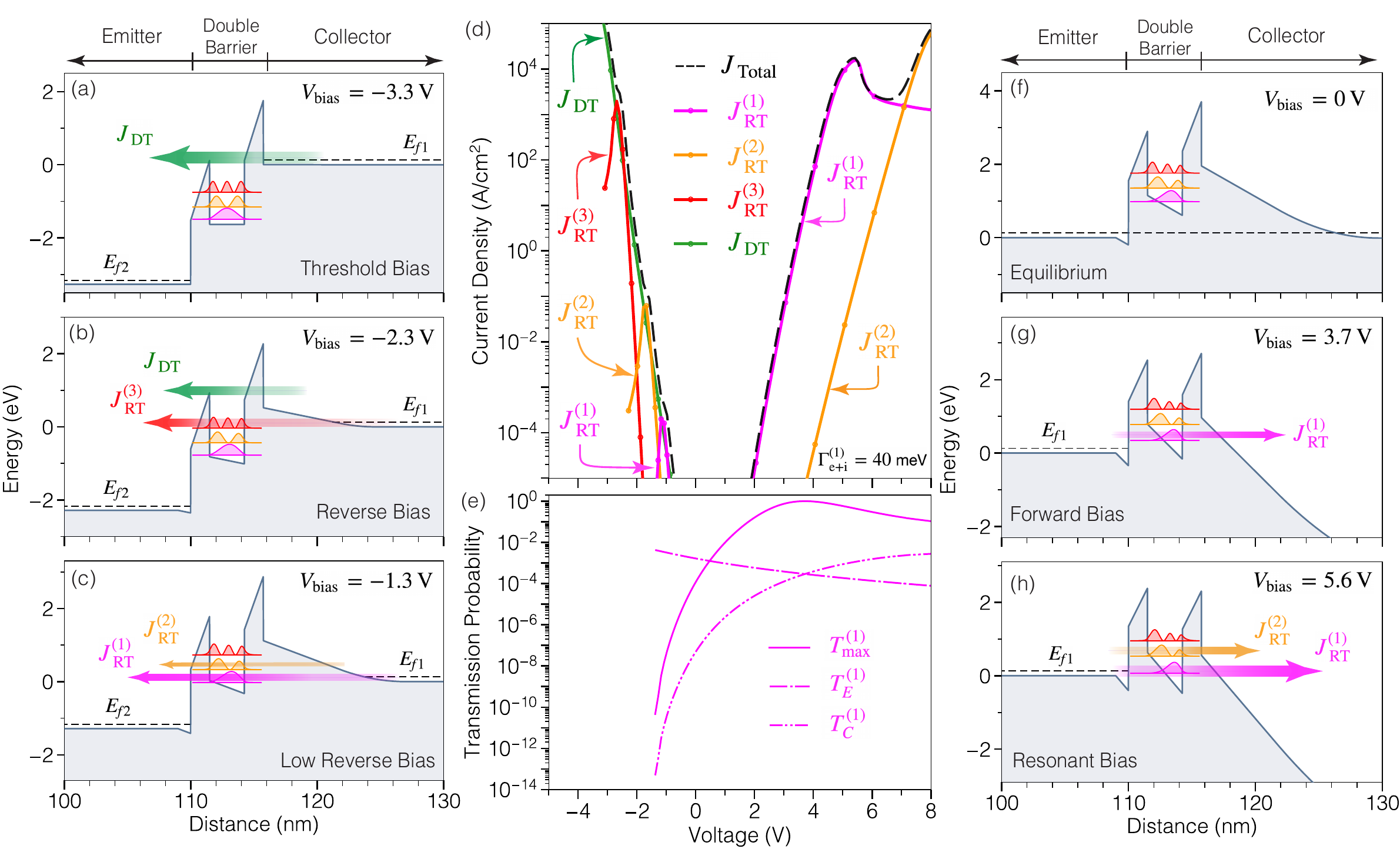}
	\caption{Quantum transport model for resonant tunneling injection across polar double-barrier heterostructures. The current-voltage characteristics of a GaN/AlN RTD are calculated using analytical expressions for the various tunneling current components. Panels (a)--(c) display the RTD band diagrams under reverse bias. (a) When the device is biased at the threshold voltage $V_\text{th}=-2t_bF_\pi$, the flatband configuration enhances the direct tunneling current component $J_\text{DT}$, which becomes the dominant transport mechanism. At this critical voltage, both AlN barriers sustain a finite electric field, which originates from the internal polarization charges. (b), (c) For voltages below the critical threshold bias (i.e. $V_\text{th}<V_\text{bias}<0$), transport is also mediated by resonant tunneling injection. In this regime, the large asymmetry between the tunneling barriers, results in strongly attenuated resonant tunneling current components ($J_\text{RT}^{(n)}$). Panel (d) displays each of these tunneling-current components. Their sum, the total tunneling current $J_\text{Total}$, is also displayed by the black dashed line. (e) As a result of the internal polarization fields, polar RTDs exhibit a strong asymmetry between the single-barrier transmission probabilities. Under reverse bias the increasing asymmetry between $T_E^{(1)}$ and $T_C^{(1)}$ leads to a highly attenuated resonant tunneling current [See $J_\text{RT}^{(1)}$ in (d)]. In contrast, under forward bias, $T_E^{(1)}$ and $T_C^{(1)}$ become more symmetric resulting in an enhancement of the maximum electron transmission $T_\text{max}^{(1)}$. Panels (f)--(h) show the RTD band diagrams at equilibrium and forward bias. (f) Under equilibrium conditions, a wide depletion region builds up next to the collector barrier and the ground-state energy rises approximately $1$~eV above the equilibrium Fermi level (black dashed line). (g) Under forward bias, the current is supported by resonant tunneling electrons injected across the double barrier structure ($J_\text{RT}^{(1)}$). The transparency of the active region increases as the tunneling transmission of each barrier becomes more symmetric. (h) Maximum resonant injection is reached at the resonant bias, when the ground-state eigenenergy aligns with the Fermi energy of the emitter reservoir. The resonant tunneling component $J_\text{RT}^{(2)}$, injected through the first excited state, is also calculated.}
\end{figure*}

\section{II. G\lowercase{a}N/A\lowercase{l}N Polar Resonant Tunneling Diodes}

To systematically investigate resonant tunneling injection across polar heterostructures, we prepare a set of double-barrier GaN/AlN RTDs. The device structures are grown by molecular beam epitaxy (MBE) on the c-plane of single-crystal GaN substrates. The growth conditions were optimized to promote step-flow growth mode, therefore minimizing the roughness of the epitaxial layers. These conditions are critical for attaining atomically smooth heterointerfaces which promote quantum interference in the tunneling electrons. The structures presented here are grown under metal-rich conditions, with the use of a constant nitrogen plasma power of 200~W and a substrate temperature fixed at 740~$^\circ$C.

Growth rates are measured $\textit{in situ}$, from the intensity oscillations of the reflection high-energy electron diffraction (RHEED) pattern. With this technique, we obtain an incorporation rate of $\sim$~7~seconds per monolayer. This result was verified, $\textit{ex situ}$, using high-resolution x-ray diffraction, which not only confirmed the growth rate but also allows us to measure the thickness and composition of the different epitaxial layers.

The inset of Fig.~1(a) shows schematically the complete epitaxial structure of the grown samples. Two degenerately doped n-GaN layers are grown as contact regions, extending $\sim 100$~nm. Silicon donors are incorporated into these layers with a nominal concentration of $\sim 2\times 10^{19}$~cm$^{-3}$. To minimize dopant diffusion into the active region, spacers are introduced next to each tunneling barrier. The thicknesses of the emitter and collector spacers are 10 and 6~nm, respectively. The double-barrier active region consists of two symmetric AlN tunneling barriers with thickness $t_b$~nm that confine the resonant states of a 3-nm-wide GaN quantum well.

Three different heterostructures with differing AlN barrier thickness ($t_b=1.5$, $2.0$, $2.4$~nm) are grown epitaxially. Control over the layer thickness at the monolayer level is achieved by tracking the RHEED intensity during the incorporation of the AlN layers. With this technique, we are able to count, in real time, the number of MLs incorporated into each tunneling barrier [\onlinecite{Tsuchiya1986}].

Figure~1(a) displays the equilibrium band diagram of each of the grown structures, calculated using a self-consistent Schr\"odinger-Poisson solver, within the framework of the effective mass equation [\onlinecite{Tan1990}]. The spontaneous and piezoelectric polarization dipoles are considered across the whole device structure. The periodic arrangement of these polarization charges results a zero average charge density within the bulk regions. In contrast, at the GaN/AlN (AlN/GaN) interfaces the intensity of the polarization dipoles exhibit a large discontinuity, which results in a net nonzero polarization charge $+q\sigma_\pi$ ($-q\sigma_\pi$). These effective polarization charges induce a redistribution of free carriers that significantly alter the conduction band profile [\onlinecite{Berland2011}]. The two-dimensional electron gas next to the emitter barrier is a consequence of the positive net polarization charge at the GaN/AlN interface ($\sigma_\pi= 6.5\times 10^{13}$~cm$^{-2}$) and its large conduction band offset, which attract and confine free carriers [\onlinecite{Ambacher2000}].

On the collector side, a similar effect occurs at the AlN/GaN interface; however the \color{m} net negative polarization charge \color{k} repels free carriers, inducing a wide depletion region [See Fig~1(a)]. The energy drop across the depleted collector can be a large fraction of the GaN bandgap ($E_g=3.4$~eV) and depends strongly on the thickness of the tunneling barriers. This effect is a direct consequence of the electric polarization of the AlN layers which sustain intense electric fields of the order of $F_b\sim 8 \text{ MV/cm}$. As a result, the effective tunneling path across both barriers is heavily modified; carriers will also tunnel across the GaN layers when the devices are biased at resonance [See Fig.~2(g)]. In addition to the strong modulation in electron transmission, thicker barriers will also generate higher ground-state energies. Self-consistent calculations indicate an energy change of approximately $100$~meV per AlN monolayer, assuming symmetric barriers. This energy shift, can be seen in Fig.~1(a) which displays the square moduli of the ground-state wavefunctions, plotted at their respective eigenenergies. Consequently, the resonant peak is expected to occur at larger peak voltages for devices with thicker tunneling barriers [\onlinecite{Encomendero2017}].

To corroborate the previous predictions, diodes were fabricated using conventional photolithography, e-beam metal evaporation, and dry etching. Details on the fabrication steps can be found in reference \onlinecite{Encomendero2017}. After processing, the devices are tested at room temperature, with injection currents up to 30~kA/cm$^2$. Typical $\textit{J-V}$ curves are presented in Fig.~1(b) and (c). Resonant tunneling injection with its hallmark NDC, is observed under forward bias in each of the fabricated devices.

As a result of the exponential relationship between barrier thickness and electron transmission, the intensity of tunneling current is modulated over several orders of magnitude. This is experimentally observed in our devices as shown in Figs.~1(b) and (c), and more clearly in Fig.~3(a). By decreasing the thickness of the AlN barriers from 2.4~nm to 1.5~nm, the peak current density increases from $2.52\times 10^2$~A/cm$^2$ to $2.55\times 10^4$~A/cm$^2$ [See Figs.~1(b) and~3(a)]. To facilitate direct comparison, in Fig.~1(b) the forward-bias current of the diodes with $t_b=2.0$~nm and $t_b=2.4$~nm is scaled by a factor of $20$ and $40$, respectively. From this figure, we also extract a maximum peak-to-valley current ratio (PVCR) of $\sim 1.55$, measured in the sample with the thinnest barriers. We attribute this enhanced PVCR to the low peak voltage, $V_p=5.66$~V, which limits the injection of excess current across the active region.

Because of the intense electric fields across the barriers and quantum well, the peak voltage shifts towards higher values for thicker barriers [See Fig.~1(b)]. The dependence of this characteristic voltage as a function of the different layer thicknesses and intensity of the polarization fields was derived previously [\onlinecite{Encomendero2017}]. Using this theory, we found that the theoretical peak voltages are consistent with the experimental values measured in our devices. 

The strong polarization fields exert an even more dramatic influence on the reverse bias injection of polar RTDs. As the reverse bias increases, the collector depletion region shrinks and the emitter barrier shifts towards lower energies with respect to the collector side, which now injects the tunneling carriers [See Fig.~3(b)]. A critical threshold voltage is reached when all the space charge within the doped layers is screened, and the depletion region is completely removed. At this bias point, the sole sources of electric field in the active region are the net polarization charges: $\pm \sigma_\pi$. The band diagram in Fig.~2(a) depicts this critical condition which depends only on two parameters [\onlinecite{Encomendero2017}]: (a) a material parameter: the magnitude of the spontaneous and piezoelectric polarization fields $F_\pi=e\sigma_\pi/\epsilon_s$, and (b) a structural parameter: the thickness of the tunneling barriers, $t_b$. This intimate connection between the critical threshold voltage and the polarization charges has been recently exploited as a sensitive measurement probe of the intense polarization fields present in III-nitride crystals [\onlinecite{Encomendero2017}]. 

The polarization-induced threshold voltage $V_\text{th}$ is also extracted from the set of devices under investigation. Figure~1(c) shows the current-voltage characteristics of these devices over a voltage range spanning from $V_\text{bias}=-6$~V up to $V_\text{bias}=10$~V. As expected, the threshold voltage exhibits a clear modulation as a function of barrier thickness. To experimentally extract $V_\text{th}$, it is important to uncouple the voltage dropped in the parasitic series resistance $\rho_s$ from the intrinsic voltage applied across the double-barrier active region [See inset in Fig.~3(c)]. To do so, we extract $V_\text{th}$ as the voltage intercept of the linear interpolation of the $\textit{J-V}$ curve under high current injection conditions (i.e. $J> 10$~kA/cm$^2$). This method is illustrated in Fig.~1(c) which also includes---in dashed lines---the linear fits to the current-voltage characteristics. The extracted $V_\text{th}$ values, exhibit a linear dependence on barrier thickness, which confirms the validity of the electrostatic model introduced previously [\onlinecite{Encomendero2017}].

After developing an analytical transport model in the next two sections, we will show that under high injection levels (i.e. $J> 10$~kA/cm$^2$), the current and voltage are both controlled mainly by the series resistance (See section V). Consequently, the linear interpolation method introduced here, allow us to accurately extract the magnitude of the threshold voltage by uncoupling the parasitic effects of the series resistance.

\begin{figure*}[t]
	\centering
	\hypertarget{Fig3}{}
	\label{Fig3}
	\includegraphics[width=\textwidth]{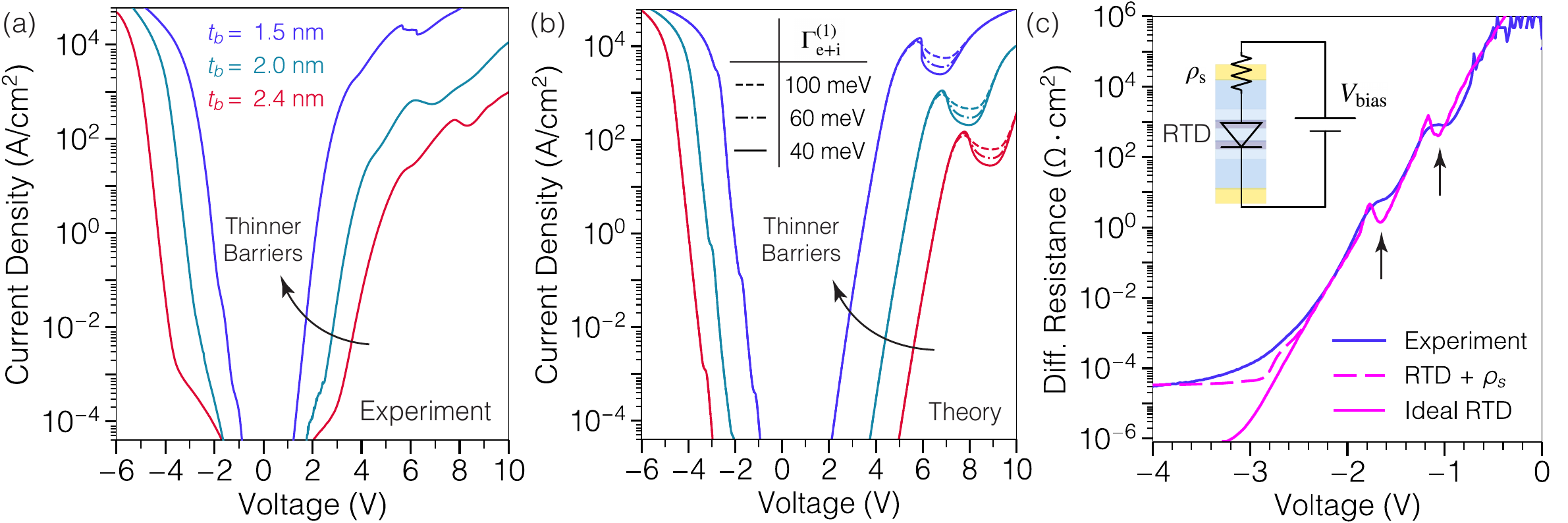}
	\caption{Comparison between experimental results and theoretical calculations obtained with our analytical resonant tunneling transport model. (a) Room-temperature $\textit{J-V}$ characteristics as a function of the barrier thickness measured from the fabricated RTDs. (b) Theoretical $\textit{J-V}$ characteristics calculated using Eq.~(\ref{eq_J_RT_final}) for different levels of incoherent tunneling transport, determined by the total width $\Gamma_\text{e+i}$. A parasitic resistance $\rho_s$, is also included in series with the ideal RTD to capture the effects of the access and contact resistances [See the inset of Fig.~3(c)]. The values of $\rho_s$ for each diode are presented in Table~(\ref{table1}). (c) Differential resistance under reverse-bias conditions for the RTD with $t_b=1.5$~nm. The arrows indicate regions of resistance modulation corresponding to the peak currents of the different resonant tunneling components. These features are completely captured by our analytical model, demonstrating a good quantitative agreement. Under high reverse bias ($V_\text{bias}<V_\text{th}$), the differential resistance is controlled mainly by the series resistance $\rho_s$. The inset shows the equivalent-circuit model of the RTD including the parasitic series resistance.}
\end{figure*}

\section{III. Coherent Tunneling Transport Model for Polar RTDs}

To quantitatively explain the experimentally measured magnitudes and shapes of resonant tunneling currents across polar heterostructures, we develop an analytical transport theory based on the quantum scattering approach introduced by Landauer and B\"uttiker [\onlinecite{Buttiker1985,Landauer1987,Buttiker1988}]. Our initial model assumes completely coherent tunneling injection. We then relax this constraint to include the effects of inelastic scattering events which are pervasive in resonant tunneling devices (See Section IV).

The expression for the tunneling current leads to the well-known Landauer formula for the case of a 1D system [\onlinecite{Buttiker1985,Landauer1987}] and to the Tsu-Esaki formula for the case of a 3D conductor [\onlinecite{Tsu1973}]:
\begin{equation}
\label{eq_Tsu-Esaki}
J=\frac{em^*kT}{2\pi^2\hbar^3}\int_0^\infty\mathcal{T}\left(E\right)\text{ln}\left[\frac{1+e^{\frac{E_{f1}-E}{kT}}}{1+e^{\frac{E_{f2}-E}{kT}}}\right]dE,
\end{equation}
where $e$, $m^*$, and $T$ are the electron charge, the effective mass and temperature, respectively; $E_{f1}$ and $E_{f2}$ are the Fermi energies of the contact regions, which supply and gather the tunneling carriers as shown in Fig.~2; $E$ is the energy of the incident electrons, $k$ and $\hbar$ are the Boltzmann and reduced Planck constants. $\mathcal{T}(E)$ is the transmission probability across the double-barrier structure, which exhibits resonances due to quantum-interference effects. Around the resonant energies $E_\text{r}^{(n)}$ ($n\in\{1,2,\cdots,N\}$), the transmission function can be approximated---to the first order---by the Breit-Wigner expression [\onlinecite{Stone1985,Price1986,Jonson1987,Kiehl1994}]:
\begin{equation}
\label{eq_T(E)}
\mathcal{T}^{(n)}(E)= T^{(n)}_\text{max}\frac{\left(\frac{\Gamma_\text{e}^{(n)}}{2}\right)^2}{\left(E-E^{(n)}_\text{r}\right)^2+\left(\frac{\Gamma_\text{e}^{(n)}}{2}\right)^2},
\end{equation}
where $\Gamma_e^{(n)}$ and $T^{(n)}_\text{max}$ are the elastic width and amplitude of the $n$th resonant peak.

The parameters of each resonant lineshape (i.e. $E_\text{r}^{(n)}$, $\Gamma_\text{e}^{(n)}$, and $T^{(n)}_\text{max}$) are bias dependent. Expressions for the elastic width and resonance amplitude can be found elsewhere [\onlinecite{Encomendero2017,Jonson1987,Kiehl1994}], however we reproduce them here for the sake of completeness.
\begin{subequations}
	\label{eq_Gamma_Tmax}
	\begin{gather}
	\frac{\Gamma_\text{e}^{(n)}}{2}=\frac{\hbar}{2}\frac{\sqrt{2E_\text{r}^{(n)}/m^*}}{2t_w}\left(T^{(n)}_E+T^{(n)}_C\right),\\
	T_\text{max}^{(n)}=\frac{4T^{(n)}_ET^{(n)}_C}{(T^{(n)}_E+T^{(n)}_C)^2},
	\end{gather}
\end{subequations}
where $t_w$ and $m^*$ are the width and effective mass of the quantum well, respectively, and  $T^{(n)}_E$ and $T^{(n)}_C$ are the emitter and collector single-barrier transmission probabilities at the $n$th resonant energy $E_\text{r}^{(n)}$. From these equations, it can be seen that the bias dependence of each resonance is captured by the following three parameters only: $E_\text{r}^{(n)}$, $T_E^{(n)}$, and $T_C^{(n)}$. These variables can be calculated from the RTD band diagram under different biasing conditions as can be seen in Figs.~2(a)-(c) and 2(f)-(h).

An important difference between polar and non-polar RTDs, is the considerable asymmetry between the effective tunneling barriers. This effect, caused mainly by the wide collector depletion region, can be seen in the equilibrium band diagram displayed in Fig.~2(f). To quantify this effect, we employ the Wentzel-Kramers-Brillouin (WKB) approximation for the calculation of the single-barrier transmission coefficients. Figure~2(e) shows $T_E^{(1)}$, $T_C^{(1)}$, and $T_\text{max}^{(1)}$ as a function of the applied voltage $V_\text{bias}$. At equilibrium, the significant asymmetry between the barrier transparencies ($T_E^{(1)}/T_C^{(1)}\sim 10^4$), strongly influences the resonance width and maximum resonant transmission of the ground state [See Eq.~(\ref{eq_Gamma_Tmax})]. However, under nonequilibrium conditions the single-barrier transmission probabilities are modulated exponentially as can be seen in Fig.~2(e).

Under forward bias, the electric field in the depleted GaN region increases, leading to an exponential increase in the collector barrier transmission as can be seen in Figs.~2(e) and (g). In contrast, the transparency of the emitter barrier decreases due to the quantum confined Stark effect (QCSE), which shifts the bound-state energy towards lower values, thus increasing the effective emitter tunneling distance. These trends make the single-barrier transmission coefficients more symmetric, thus enhancing the double-barrier resonant transmission $T_\text{max}^{(1)})$ [See Fig.~2(e)].

It is well known that perfect transparency of the double-barrier structure (i.e. $T^{(n)}_\text{max}=1$) is attained only when both barriers exhibit equal transmission coefficients: $T^{(n)}_E=T^{(n)}_C$ [\onlinecite{Ricco1984}]. For the case of non-polar RTDs with symmetric barriers, prefect transmission occurs at equilibrium, without carrier injection. In contrast, polar RTDs are able to reach perfect resonant transparency under non-equilibrium regime, enabling maximum carrier injection into the resonant levels.

For the GaN/AlN RTD presented in Fig.~2, symmetric single-barrier transparency (i.e $T_E^{(1)}=T_C^{(1)}$) and unitary resonant transmission (i.e. $T_\text{max}^{(1)}=1$) is attained when $V_\text{bias}=3.7$~V, as can be seen in Figs.~2(e) and (g). At this point, the active region is completely transparent to electrons injected at the eigenenergy $E_r^{(1)}$. However, since the resonant level is still $\sim 300$~meV above the emitter Fermi level, carrier supply is limited by the Fermi-Dirac tail at the emitter contact. Consequently, if the forward bias is increased, a larger number carriers will be available for conduction and the current will increase until the subband energy aligns with the Fermi level $E_{f1}$. This resonant configuration occurs at the peak voltage $V_\text{bias}=5.6$~V [See Fig.2(h)], with a resonant transmission peak $T_\text{max}^{(1)}\sim 10^{-1}$, as can be seen from Fig.~2(e). From this analysis, it is clear that higher peak current levels are expected from polar double-barrier structures designed to exhibit perfect carrier transparency right at the resonant voltage.

Under reverse bias, however, perfect transparency across the double-barrier structure cannot be attained for the case of polar RTDs with symmetric barriers. This can be seen from Figs.~2(b) and (c) which show the device band diagrams under reverse bias conditions, and from Fig.~2(e) which shows the calculated transmission probabilities. As the applied voltage becomes more negative, the transparency of the collector barrier decreases since the quasibound state shifts towards lower energies. The net result of this transition is an increase in the effective tunneling distance, leading to a more opaque collector barrier at the resonant energies [See Fig.~2(c)]. In contrast, the transparency of the emitter barrier increases rapidly due to the larger intensity of the electric field across the AlN barriers [See Fig.~2(b)]. These opposite trends in single-barrier transmission coefficients result in a heavily attenuated tunneling injection under reverse bias. This can be seen in Fig.~2(d) which displays the calculated resonant tunneling currents $J_\text{RT}^{(n)}$ injected through the $n$th resonant level. These results are discussed in the next section after presenting an analytical expression for the resonant tunneling current components.

As discussed in section II, a critical flatband configuration is reached when the width of the depletion region is completely suppressed. From Fig.~2(a), it is clear that the tunneling current at this point is not supported by resonant tunneling injection since the quasi-bound states are well below the energies at which the collector supplies tunneling carriers. Instead, electrons are injected by direct tunneling through the collector barrier, which sustains an electric field equivalent to the internal polarization field ($F_\pi$). The direct tunneling current $J_\text{DT}$, can be calculated using the Tsu-Esaki integral and the tunneling transmission function across the effective collector barrier $\mathcal{T}_\textit{Col.}(E)$:
\begin{equation}
\label{eq_J_DT}
J_\text{DT}=\frac{em^*kT}{2\pi^2\hbar^3}\int_0^\infty\mathcal{T}_\textit{Col.}\left(E\right)\text{ln}\left[\frac{1+e^{\frac{E_{f1}-E}{kT}}}{1+e^{\frac{E_{f2}-E}{kT}}}\right]dE,
\end{equation}
$E_{f1}$ and $E_{f2}$ are the Fermi levels of the reservoirs that supply and gather the tunneling carriers as labeled in Fig.~2(a). $\mathcal{T_\textit{Col.}}(E)$ is calculated using the WKB approximation which can be employed for an arbitrary tunneling barrier:
\begin{equation}
\label{WKB_T_col}
\mathcal{T}_\textit{Col.}\left(E\right)\approx \exp\left(-2\int_\varnothing \sqrt{2m^*\left(x\right)\left[E_c(x)-E\right]}/\hbar\,dx\right),
\end{equation}
where $\varnothing$ represents the tunneling path across the effective potential barrier. $E_c\left(x\right)$ is the conduction band profile and $m^*\left(x\right)$ is the effective mass, which is position dependent along the integration path $\varnothing$. Figure~2(d), shows the calculated direct tunneling current $J_\text{DT}$ which exhibits a monotonic increment as the collector barrier becomes increasingly transparent to the incident electrons.

\section{IV. Coherent and Sequential Tunneling Transport Model for Polar RTDs}

In this section, we consider the consequences of inelastic scattering events on the resonant tunneling current injected across polar heterostructures. In contrast to the completely coherent tunneling picture discussed in the previous section, the presence of inelastic scatterers leads to the loss of phase coherence in a subset of the tunneling electrons. As a result, scattered carriers traverse the active region in a two-step tunneling process known as ``sequential tunneling" [\onlinecite{Luryi1985,Buttiker1988}]. In this inelastic transport regime, electrons that lose phase memory give rise to an incoherent current component which enhances off-resonant transmission.

The incoherent contribution to the current can be calculated employing Landauer's transmission theory of quantum transport. B\"uttiker obtained analytical expressions for the coherent and incoherent transmission probabilities across a general double-barrier structure. From this analysis, the total probability of transmission through the $n$th resonant level can be expressed by the sum of the coherent and incoherent probabilities [\onlinecite{Buttiker1988}]:
\begin{equation}
\label{eq_T_coh_incoh}
\mathcal{T}_{{\scriptstyle\text{Coh.+}}\atop{\scriptstyle\text{Incoh.}}}^{(n)}(E)= T^{(n)}_\text{max}\left(\frac{\Gamma_\text{e}^{(n)}}{\Gamma_\text{e+i}^{(n)}}\right)\frac{\left(\frac{\Gamma_\text{e+i}^{(n)}}{2}\right)^2}{\left(E-E^{(n)}_r\right)^2+\left(\frac{\Gamma_\text{e+i}^{(n)}}{2}\right)^2},
\end{equation}
where $T^{(n)}_\text{max}$, $\Gamma_\text{e}^{(n)}$, and $E_r^{(n)}$ were defined previously [See Eq.~(\ref{eq_Gamma_Tmax})] and $\Gamma_\text{e+i}^{(n)}$ is the total transmission width of the $n$th resonant level given by the sum of the elastic ($\Gamma_\text{e}^{(n)}$) and inelastic ($\Gamma_\text{j}^{(n)}$) partial widths [\onlinecite{Buttiker1988}].

Equation~(\ref{eq_T_coh_incoh}) reveals that phase-randomization events result in the attenuation of the resonant transmission peak by a factor of $\Gamma_\text{e}^{(n)}/\Gamma_\text{e+i}^{(n)}$, and a concomitant broadening of the resonance linewidth [\onlinecite{Jonson1987,Buttiker1988}]. Scattering processes include optical phonons [\onlinecite{Goldman1987}], intrasubband and intersubband transitions [\onlinecite{Grier2015}], and interface roughness scattering events [\onlinecite{Song2016}]. In the case of GaN/AlGaN heterostructures, interface roughness scattering and electron-phonon interactions have been identified as the main processes leading to decoherence [\onlinecite{Grier2015,Song2016}].

In the present discussion, we focus on the overall effects of the different scattering mechanisms, which we assume to be uncorrelated. In this case, the total resonance width $\Gamma_\text{e+i}^{(n)}$ can be expressed by the sum of the different elastic and inelastic partial widths: $\Gamma^{(n)}_\text{e+i}=\Gamma^{(n)}_\text{e}+\sum_\text{j}\Gamma^{(n)}_\text{j}$, with $\Gamma_\text{j}^{(n)}$ being the $j$th inelastic partial width [\onlinecite{Buttiker1988}]. Each scattering mechanism is characterized by a dephasing time $\tau^{(n)}_\text{j}$, which determines its corresponding inelastic width. Meanwhile the dwell time inside the resonant tunneling region, $\tau^{(n)}_\text{e}$, determines the RTD elastic width. Using the Planck-like expression $\Gamma^{(n)}_\text{j}=\hbar/\tau^{(n)}_\text{j}$ ($\Gamma^{(n)}_\text{e}=\hbar/\tau^{(n)}_\text{e}$), one can obtain an effective scattering time: $1/\tau_\text{e+i}^{(n)}=1/\tau_\text{e}^{(n)}+\sum_j1/\tau_\text{j}^{(n)}$. The relation between the total width and the effective scattering time, given by $\Gamma^{(n)}_\text{e+i}=\hbar/\tau^{(n)}_\text{e+i}$, allows us to use the total width as a phenomenological parameter which characterizes the mixed---coherent and incoherent---tunneling transport across the polar double-barrier heterostructure. Typical values for this parameter are introduced in the next section and its effects on the RTD current-voltage characteristics and PVCR are also discussed.

An analytical expression for the resonant tunneling current $J_\text{RT}^{(n)}$, which includes contributions from coherent and sequential tunneling processes, is derived by replacing the Breit-Wigner form of the resonance transmission [Eq.~(\ref{eq_T_coh_incoh})] in the Tsu-Esaki integral [Eq.~(\ref{eq_Tsu-Esaki})]. The supply function, given by the logarithmic factor inside the integrand in Eq.~(\ref{eq_Tsu-Esaki}), is a slowly varying function compared to the resonant factor $\mathcal{T}_{{\scriptstyle\text{Coh.+}}\atop{\scriptstyle\text{Incoh.}}}^{(n)}(E)$, which exhibits a narrow transmission peak. Consequently, the main contribution to the tunneling current is provided by electrons with injection energies around $E_\text{r}^{(n)}$. If the supplied carriers exhibit an energy range that is much larger than the total resonance width, we can approximate the supply function as a constant for energies close to the subband energy. This condition is satisfied when $E_{f1}\gg \Gamma^{(n)}_\text{e+i}$, which is usually the case when degenerately doped layers are employed as carrier injection contacts. As a result, the supply function can be taken out of the integral with the substitution $E=E_\text{r}^{(n)}$ [\onlinecite{Schulman1996}]. Finally, after integrating the Lorentzian function, we can write an analytical expression for the resonant tunneling current injected through the $n$th quasibound state:
\begin{widetext}
	\begin{equation}
	\label{eq_J_RT_final}
	J^{(n)}_{\text{RT}}\left({\scriptstyle V_\text{bias}}\right)=\frac{qm^*kT}{2\pi^2\hbar^3}T^{(n)}_\text{max}\left({\scriptstyle V_\text{bias}}\right)\left[\frac{\Gamma_e^{(n)}\left({\scriptstyle V_\text{bias}}\right)}{2}\right]\text{ln}\left[\frac{1+\exp\left(\frac{E_{f1}\left({\scriptscriptstyle V_\text{bias}}\right)-E^{(n)}_\text{r}\left({\scriptscriptstyle V_\text{bias}}\right)}{kT}\right)}{1+\exp\left(\frac{E_{f2}\left({\scriptscriptstyle V_\text{bias}}\right)-E^{(n)}_\text{r}\left({\scriptscriptstyle V_\text{bias}}\right)}{kT}\right)}\right]\left[\text{tan}^{-1}\left(\frac{{\scriptstyle E^{(n)}_\text{r}\left({\scriptscriptstyle V_\text{bias}}\right)}}{\frac{\Gamma_\text{e+i}^{(n)}\left({\scriptstyle V_\text{bias}}\right)}{2}}\right)+\frac{\pi}{2}\right].
	\end{equation}
\end{widetext}

Using Eq.~(\ref{eq_J_RT_final}), we calculate the different resonant tunneling current components for a GaN/AlN RTD, whose device structure is shown in Fig.~1(a), with $t_b=1.5$~nm. Figure~2(d) displays the current-voltage curves $J_\text{RT}^{(n)}$, calculated for each of the quasibound states localized in the double-barrier active region. The single-barrier transmission coefficients $T_E^{(n)}({\scriptstyle V_\text{bias}})$ and $T_C^{(n)}({\scriptstyle V_\text{bias}})$ are computed using the WKB expression given by Eq.~(\ref{WKB_T_col}); $\Gamma_\text{e}^{(n)}({\scriptstyle V_\text{bias}})$ and $T_\text{max}^{(n)}({\scriptstyle V_\text{bias}})$ are obtained from Eq.~(\ref{eq_Gamma_Tmax}). The bias dependence of these parameters is contained in the RTD band diagrams which are calculated at every bias point [See Figs.~2(a)-(c) and (f)-(h)].

The RTD band diagrams are calculated with the use of the analytical model for polar RTDs introduced previously by the authors [\onlinecite{Encomendero2017}]. This approximate model takes into consideration the space-charge regions, the free-carrier distribution, and the net spontaneous and piezoelectric polarization charges induced at the polar heterointerfaces, as discussed in section II. Under close-to-equilibrium conditions the resonant levels are completely unpopulated, as can be seen from Fig.~2(f). However under moderate bias, carriers tunnel into the well, thereby increasing the subband population and building up the charge inside the well. To calculate the subband population under non-equilibrium conditions, a numerical self-consistent procedure is usually used to solve the Poisson and Schr\"odinger equations simultaneously [\onlinecite{Cahay1987,Klimeck1995}]. However, for the present discussion it can be shown that the analytical form of the RTD band diagram provides a reasonable approximation. To do so, we consider the dwell time $\tau_e^{(1)}\approx 7$~ps, calculated for the case of the RTD with $t_b=1.5$~nm. Taking into account the RTD peak tunneling current $J_p\approx 2.5\times 10^4$~A/cm$^2$, we calculate that the electron concentration inside the well---at maximum carrier injection---will be: $\sigma_w\sim 1.0\times 10^{12}$~cm$^{-2}$. Therefore, the charge accumulated inside the well is almost two orders of magnitude lower than the net polarization charges ($\sigma_\pi=6.5\times 10^{13}$~cm$^{-2}$) present at the well--barrier interfaces. After performing a similar analysis in the other RTDs, we conclude that our analytical band diagram provides an adequate approximation for the magnitudes of resonant tunneling current discussed in the present work.

As can be seen from Fig.~2(d), the resonant tunneling current transmitted through the ground-state level $J_\text{RT}^{(1)}$, exhibits broken symmetry with respect to the bias polarity. This asymmetric injection is a direct consequence of the charge segregation at equilibrium which results in a wide depletion layer on the collector side. From Fig.~2(f), we can see that the depleted region acts effectively as an additional barrier with height $\sim 1.9$~eV. The extension of this space-charge region increases with the applied voltage, thus generating a stronger electric field. As a consequence, a large portion of the biasing voltage is dropped across this depletion layer [See Fig.~2(g)]. This leaves only a small fraction of the biasing voltage to modulate the double-barrier and quantum well active region. Because of this nonuniform bias distribution, polar RTDs exhibit a limited modulation control over the quasibound state energies under forward bias. For the III-nitride RTD presented in Fig.~2, we obtain a ground-state energy modulation of $\gamma_F\sim 180$~meV/Volt under forward bias. In this case, since $E_\text{r}^{(1)}\sim 1$~eV at equilibrium, the expected peak voltage will be $V_p=5.6$~V, as can be seen from Fig.~2(d).

\begin{table}[b]
	\caption{\label{table1}
		Experimental and theoretical GaN/AlN RTD parameters. The characteristic threshold voltage ($V_\text{th}$) and parasitic series resistance ($\rho_s$) are measured by a linear-interpolation procedure, introduced in Sec. II. The experimental $V_\text{th}$ values extracted from our devices agree very well with theoretical values obtained from an analytical electrostatic model [\onlinecite{Encomendero2017}]. The interpolation method requires that when $V_\text{bias} < V_\text{th}$, the series resistance is much larger than the intrinsic RTD resistance ($\rho_\text{RTD}$) [See Fig.~3(c)]. This assumption is justified since $\rho_s\gg \rho_\text{RTD}$ for each RTD presented here.}
	\begin{ruledtabular}
		\begin{tabular}{cccc}
			\text{Barrier Thickness} (\text{nm}) & 1.5 & 2.0 &2.4\\
			\hline
			\text{Theoretical} $V_\text{th}\,$(V) & 3.3 & 4.4 & 5.3\\
			\text{Experimental} $V_\text{th}\,$(V) & 3.3 & 4.3 & 5.2\\
			$\rho_s\,$ ($\Omega\cdot$cm$^2$) & $3.0\times 10^{-5}$ & $2.3\times 10^{-5}$ & $1.7\times 10^{-5}$ \\
			$\rho_\text{RTD}$ ($\Omega\cdot$cm$^2$)\footnote{Calculated at the threshold bias ($V_\text{bias}=V_\text{th}$)}  & $8.4\times 10^{-7}$ & $8.9\times 10^{-7}$ & $9.4\times 10^{-7}$ \\
		\end{tabular}
	\end{ruledtabular}
\end{table}

Under reverse bias, the depletion region also exerts a strong influence over the modulation of the quasibound state energies. However in this regime, as free carriers gather at the edge of the depletion layer, the space-charge extension reduces. As a result, the depletion barrier is lowered due to the less-intense electric field and much narrower depletion region [See Fig.~2(b) and (c)]. This charge transfer also leads to a concomitant increase in the electric field across the tunneling barriers which push the GaN quantum well towards lower energies. Both of these effects contribute to a more effective modulation of the subband energies, resulting in a reverse bias shift of $\gamma_R\sim 770$~meV/Volt for the ground-sate of the RTD presented in Fig.~2. The stronger modulation of the quasi-bound state energies results in resonant tunneling peaks occurring within a smaller voltage range compared to the forward-bias regime. This effect, combined with the presence of a direct tunneling injection process across the collector barrier ($J_\text{DT}$), results in a total tunneling current ($J_\text{Total}=J_\text{DT}+\sum_nJ_\text{RT}^{(n)}$) in which the resonant tunneling peaks are partially masked [See Fig.~2(d)]. Experimental evidence of these resonant tunneling phenomena under reverse bias is presented in the next section in which we demonstrate the agreement between theoretical calculations and experimental measurements obtained from III-nitride RTDs.

\section{V. Discussion and Conclusions}

The analytical tunneling theory introduced in section IV is used to understand the phenomenon of resonant tunneling injection in GaN/AlN RTDs. Figure~3(a) shows the room temperature current-voltage characteristics measured from the series of RTDs presented in Fig.~1. From this semilogarithmic plot, it is evident that the AlN barriers exert an exponential control over the magnitude of tunneling current under both bias polarities. At low-bias injection, the wide depletion barrier effectively blocks carrier transport, resulting in low current densities. When the applied voltage is increased, the quasibound state energies and height of the depletion barrier are modulated, enabling carrier injection into the resonant levels. However the nonuniform distribution of the applied bias across the active region, results in asymmetric current injection as can be seen from Fig.~3(a).

For each device structure, we calculate the total tunneling current including contributions from direct and resonant tunneling injection processes. Figure~3(b) shows that the experimental results are reproduced very well by our theoretical model. Good agreement is observed over the entire voltage range, with injection currents spanning several orders of magnitude. Devices with the same barrier thickness are plotted using the same color scheme employed in Fig.~3(a). To capture the effects of the access and contact resistances, a parasitic resistor is also included in the theoretical calculation. The equivalent circuit model is shown in the inset of Fig.~3(c), comprising the ideal RTD connected in series with a resistor of magnitude $\rho_s$.

To get further insight into the reverse-bias tunneling transport, we analyze the experimental differential resistance shown in Fig.~3(c). For clarity, we only present the data for the RTD with $t_b=1.5$~nm, however a similar analysis has been done for the other devices. Under low-bias injection ($-2$~V$<V_\text{bias}$), two clear features are observed in the experimental data, indicated by the black arrows. Our model reproduces very well the occurrence of these tunneling features which correspond to peaks in the current injected through the first two subband levels via resonant tunneling [See Fig.~2(d)]. At high injection levels ($V_\text{bias}<-2$~V), there is an evident deviation between the experimental and ideal-RTD differential resistance calculated with our model. This discrepancy can be explained by the presence of a parasitic series resistance $\rho_s$ which controls the current flow at high injection levels [See inset in Fig.~3(c)].

Identifying the voltage dropped across the parasitic resistance is important to determine the intrinsic RTD voltage and its characteristic threshold voltage $V_\text{th}$. The extraction of these two parameters, $\rho_s$ and $V_\text{th}$, is done following the interpolation procedure discussed in section II. This method assumes that when $V_\text{bias}<V_\text{th}$, the series resistance is much larger than the intrinsic RTD resistance. We can verify that this is the case comparing the differential resistance of the ideal RTD ($\rho_\text{RTD}$) with the magnitude of $\rho_s$ when the device is biased at $V_\text{bias}=V_\text{th}$. These values are presented in Table~(\ref{table1}), confirming that $\rho_s\gg\rho_\text{RTD}$ for each of the RTDs presented here; thus the linear interpolation procedure is justified. Furthermore, this is also corroborated by the good quantitative agreement between theoretical and experimental values of the critical threshold voltage $V_\text{th}$ [See Table~(\ref{table1})].

Under forward bias, Fig.~3(b) shows that our analytical model correctly reproduces the dependence of the peak voltage on the barrier thickness. This modulation is completely captured by the device band diagram, from which the bias-dependent parameters in Eq.~(\ref{eq_J_RT_final}) are calculated. In particular, the onset of NDC will be determined by the detuning of the subband energy $E_r^{(n)}\left(V_\text{bias}\right)$ with respect to the conduction band minimum on the emitter side. Furthermore, the region of NDC and PVCR, depend on the total width $\Gamma_\text{i+e}$, which we consider a phenomenological parameter related to the dephasing rate inside the double-barrier structure.

Previous theoretical studies showed that typical dephasing times in GaN/Al$_x$Ga$_{1-x}$N RTDs at room temperature are of order of approximately $33$~fs, which corresponds to a total broadening of $\Gamma_\text{e+i}\approx 20$~meV [\onlinecite{Grier2015}]. On the other hand, intersubband absorption experiments performed on GaN/AlN multiple quantum wells have revealed even lower scattering times of $\sim 10$~fs ($\Gamma_\text{e+i}\approx 66$~meV). These ultrafast dephasing rates result from considerable interface roughness scattering which is shown to be dominant over the LO-phonon scattering ($\tau_\text{LO}\sim 50$~fs) [\onlinecite{Song2016}]. 

In our calculations we employ three different magnitudes of resonance broadening to illustrate the important effects of the incoherent tunneling contribution. The results presented in Fig.~3(b) show that as the broadening $\Gamma_\text{e+i}$ increases, the peak tunneling current decreases with an increasing number of carriers traversing the active region via sequential tunneling. Consequently, off-resonant transmission increases, leading to a larger valley current and reduced PVCR. In the case of the RTD with $t_b=1.5$~nm, we calculate a theoretical PVCR of 2.2 (6.7), considering a broadening of $100$~meV ($40$~meV), which corresponds to $6.6$~fs ($16.5$~fs). It should be noted however that these results consist only of coherent and sequential tunneling phenomena without taking into consideration additional leakage paths which also degrade the resulting PVCR.

In summary we have systematically investigate the phenomenon of resonant tunneling injection through polar double-barrier heterostructures. Epitaxial control at the monolayer level, is used to engineer different magnitudes of resonant tunneling current injection in GaN/AlN double-barrier heterostructures. Analysis of their current-voltage characteristics reveals the control exerted by the polar tunneling barriers over the magnitude of the injected current, broadening of the resonant tunneling line shape, and critical threshold voltage. Additionally, new tunneling features are identified under reverse bias injection, which originate from highly attenuated tunneling carriers injected via resonant tunneling transport across the quasibound states.

On the theoretical side, we introduce an analytical theory for resonant tunneling transport across polar heterostructures. A general expression for the resonant tunneling current, including contributions from coherent and sequential tunneling processes is presented. Using this model, we calculate the current-voltage characteristics of GaN/AlN resonant tunneling diodes, demonstrating a good agreement with experimental results over both bias polarities, and with tunneling currents spanning several orders of magnitude. The analytical form of our model, provides a simple expression that elucidates the connection between the double-barrier design parameters and the resulting current-voltage characteristics and tunneling transport across a general polar RTD. This new theory paves the way for the design of polar resonant tunneling devices exhibiting efficient resonant injection and enhanced tunneling dynamics as required in various practical applications.

\section{VI. Acknowledgments}

This work was funded by the Office of Naval Research under the DATE MURI Program (Contract: N00014-11-10721, Program Manager: Dr. Paul Maki) and the National Science Foundation (NSF) MRSEC program (DMR-1719875). The authors also acknowledge partial support from NSF-DMREF (DMR-1534303) and Emerging Frontiers in Research and Innovation (EFRI) NewLAW (Grant No. EFMA-1741694) programs. This work was performed in part at Cornell NanoScale and Technology Facility, an National Nanotechnology Coordinated Infrastructure member supported by NSF Grant NNCI-1542081, and Cornell Center for Materials Research Shared Facilities which are supported through the NSF MRSEC program (DMR-1719875).

\bibliography{z_refs}

\end{document}